# Causation versus Prediction: Comparing Causal Discovery and Inference with Artificial Neural Networks in Travel Mode Choice Modeling

**Rishabh Singh Chauhan**
Department of Civil, Materials, and Environmental Engineering
University of Illinois Chicago, Chicago, IL 60607
Email: rchauh6@uic.edu

**Uttara Sutradhar**
Department of Civil, Materials, and Environmental Engineering
University of Illinois Chicago, Chicago, IL 60607
Email: usutra2@uic.edu

**Anton Rozhkov**
College of Urban Planning and Public Affairs
University of Illinois Chicago, Chicago, IL 60607
Email: arozhk2@uic.edu

**Sybil Derrible**
Department of Civil, Materials, and Environmental Engineering
University of Illinois Chicago, Chicago, IL 60607
Email: derrible@uic.edu

Word Count: 6823 words + 2 table (250 words per table) = 7,323 words

## ABSTRACT

This study compares the performance of a causal and a predictive model in modeling travel mode choice in three neighborhoods in Chicago. A causal discovery algorithm and a causal inference technique were used to extract the causal relationships in the mode choice decision making process and to estimate the quantitative causal effects between the variables both directly from observational data. The model results reveal that trip distance and vehicle ownership are the direct causes of mode choice in the three neighborhoods. Artificial neural network models were estimated to predict mode choice. Their accuracy was over 70%, and the SHAP values obtained measure the importance of each variable. We find that both the causal and predictive modeling approaches are useful for the purpose they serve. We also note that the study of mode choice behavior through causal modeling is mostly unexplored, yet it could transform our understanding of the mode choice behavior. Further research is needed to realize the full potential of these techniques in modeling mode choice.

## INTRODUCTION

Data modeling is central to transportation planning. Choosing the appropriate modeling approach is critical to be able to achieve one's goals, but it is not trivial. In particular, the importance of the modeling purpose needs to be emphasized as different modeling approaches serve different purposes. As Shmuell (1) explained, modeling approaches can be broadly categorized into three main categories: explanatory modeling, predictive modeling, and descriptive modeling. Explanatory modeling tests causal hypothesis about some theoretical constructs using statistical models on data. Predictive modeling predicts new or future observations by using statistical models or data mining algorithms on data. Descriptive modeling summarizes or represents the data structure in a compact manner with the causal theory being either absent or implicit. This article focuses on the first two modeling approaches – explanatory and predictive. Explanatory or causal modeling (as we refer to it throughout this article) is divorced from predictive modeling, especially when dealing with finite quantity of real data since the optimal model for prediction purpose may be different from the one estimated to explore the underlying mechanism (2). Transportation researchers often have to choose between causation and prediction when developing a model, which can be seen as a trade-off between causation and prediction (3).

Understanding the fundamental differences between causal and predictive modeling is important. As explained in Shmuell (1), let us hypothesize that $\chi$ causes $\Upsilon$ through a function F such that $\Upsilon$ = F($\chi$). Measurable variables *X* and *Y*, and a function *f* are the operationalizations of $\chi$, $\Upsilon$, and F respectively such that $E(Y) = f(X)$. In causal modeling, the variables *X* and *Y* are used to estimate *f* with the goal of matching *f* to *F*. In contrast, in predictive modeling, *X* and *Y* directly are of interest, and *f* is used to predict the new values of *Y*. It is even possible that a function other than $\hat{f}(X)$ might be preferable for prediction despite the causal relation being $\Upsilon$ = F($\chi$). Thus, causal modeling is based on causation while predictive modeling is based on associations (or correlations). Causal modeling is retrospective while predictive modeling is prospective. Further, the accuracy of predictive modeling is easier to test and observe, while the results from causal modeling can never be fully confirmed.

Data modeling in transportation planning can have various applications like understanding travel mode choice decision making, traffic crash detection, traffic delay prediction, and understanding vehicle ownership. While predictive modeling would be more appropriate for some applications, causal modeling would be for others. For example, predictive modeling would be an obvious choice for traffic crash detection (4), but causal modeling would be more appropriate to understand carpooling behavior (5). Travel mode choice models could be used to estimate the demand for the various travel modes and examine the factors that affect mode choice. Traditionally, regression-based models, particularly the multinomial logit model (MNL), have been used for modeling mode choice (6). MNL models provide a closed-form mathematical formulation and are interpretable based on random utility (6,7). More recently, machine learning (ML) methods have found their application in mode choice modeling. ML models are known to offer high accuracy in mode choice prediction (6). Despite the advantages that statistical and ML models offer, they lack an important aspect which is causality (8). Virtually every statistical and ML models are based on correlations but not causation, and hence, these cannot be used for causal modeling (8).

The literature on correlation-based modeling or predictive modeling of mode choice is abundant. In contrast, the literature on causal modeling of mode choice is scant. An important reason for this is that determining causality from observed data is complicated. To know whether a certain variable causes another variable or not, we must compare the outcomes *with* versus *without* the presence of that variable. Unfortunately, in the real world, only one of the outcomes can be observed. This leads to the fundamental problem of causal inference (9). To circumvent the major challenges associated with determining causality, researchers have developed specialized causal models to address causality. The two major techniques to study causality are causal discovery and causal inference. Causal discovery is the process of extracting causal relationships directly from observational data, while causal inference is the process of estimating the quantitative causal effects from a change of a certain variable (cause) over an outcome of interest (effect) (10). Despite being in their infancy, these emerging methods have found applications in several disciplines.

The goal of this study is threefold: (a) to develop causal models for mode choice using causal discovery and inference; (b) to compare the performance of causal models with predictive models and to highlight the key differences in the analysis from each of the techniques; and (c) to better understand the mode choice behavior in Chicago neighborhoods using causal as well as predictive models.

## LITERATURE REVIEW

Travel behavior modeling has been integral to transportation planning since its earliest applications in cities like Chicago and Detroit several decades ago (11). Mode choice modeling is a one of the most classic problems in transportation studies (12,13). Several methods have been proposed in the literature to perform mode choice modeling. Discrete choice models based on random utility theory are one of the most extensively used models for mode choice modeling (14). Discrete choice models predict the choice between two or more available alternates that are mutually exclusive and collectively exhaustive (15). Random utility models assume that the decision maker selects the alternative that offers the highest utility (16). The examples for discrete choice models based on random utility theory include MNL and probit models. Several studies have used these models to study mode choice (6,17,18).

In recent years, ML has been successfully applied to several mode choice modeling problems. Some of the ML algorithms that have been used to model mode choice include random forest (20), support vector machines (21), extreme gradient boosting (7), and artificial neural networks (ANNs) (6,22). Several studies have compared the performances of the various ML algorithms for mode choice prediction, albeit with no final consensus (22–26), while some others found ANNs to perform best (23,24). ANNs have been praised for their superior prediction performance and the ability to capture non-linear relations (6,23,27).

Statistical models, like MNL, offer good interpretability and are used to understand the underlying mechanism. However, since they are based on correlations, their findings cannot be interpreted in terms of causality. Similarly, despite all the advantages that ML algorithms offer, they are appropriate only for prediction or correlation-based interpretation. These techniques have not been able to provide causal understanding. Regardless, there has been a keen interest to unravel causality in mode choice behavior (28–34). One of the methods that researchers have used to estimate causal models for mode choice are structural equation models (SEM) (28,34). SEMs are considered to be causal models but they are confirmatory tools, not exploratory (35). The accuracy of their causal findings depends upon the accuracy of the causal assumptions made by the researchers (35). SEMs take these causal assumptions as their inputs and test their suitability on data. They are not useful in extracting causal relations directly from data (36,37). It is a major limitation of most of the causal models used for mode choice that these studies rely upon the hypothetical causal structures assumed by researchers (28,31,33,34).

As of this writing, only a handful of studies derive an underlying causal network behind the mode choice decision making process directly from data (36,38–40). These studies have applied different algorithms to study causality. Our study aims to extend the research on causal models for mode choice to new scenarios and datasets.

## METHODS

### Data Preparation

The data used in this study were collected by My Daily Travel survey conducted by the Chicago Metropolitan Agency for Planning (CMAP) (41). The survey was conducted in northeastern Illinois (including the city of Chicago) and was completed in 2019. The survey asked the respondents about their travel behavior and choices. Since the travel behavior within the city might be quite different than that in surrounding areas, we only focused on the data collected from Chicago. Based on our domain knowledge and available literature, we filtered out several variables that might be relevant to mode choice. These variables can be categorized into three categories: socio-demographic information, trip characteristics, and mode choice. They are listed in Table 1.

The survey data were first cleaned. Any invalid or unknown responses were removed. These include responses like ‘not ascertained’, ‘I don’t know’, ‘I prefer not to answer’, and appropriate skip. To narrow the scope of the study, only private vehicles (car), public transit, and walking modes were studied, while trips with other modes were filtered out. The variables were converted to ordinal or binary variables before modeling. To have multiple datasets to study the model performances and to capture the regional variation in travel behavior, the data from Chicago was spilt into three datasets each corresponding to the three neighbourhoods in the city.

Dividing the city into these macro clusters recognizes the unique characteristics and challenges faced by different regions. This is particularly important for Chicago since it is defined as a "city of neighborhoods" with stark socioeconomic disparities critically fluctuating from neighborhood to neighborhood due to the historical context of neoliberal governance, with exclusion, segregation, and redlining hidden under the policies (42). These disparities often align with geographical locations, creating distinct pockets of affluence as well as disadvantages and disinvested areas (43,44). To address these issues, splitting Chicago into neighborhoods is important for our analysis as well as for policy recommendations regarding the allocation of resources, infrastructure development, and public services.

We used Census data (45) at the standard division of the level of 77 community areas which the city of Chicago uses to collect data (46,47). These areas were grouped by socio-economical similarities into 3 bigger clusters, namely North side, West side, and South side. At this geographical level, each of these neighborhoods represents their unique socioeconomic dynamics. The North side and the central part (including Loop – central business district) is predominantly characterized by higher-income neighborhoods with significant white and Asian populations and robust commercial and business-centric nature. The South side experiences greater poverty rates, higher unemployment, transit and food deserts, and a larger Black population. Finally, the West side is somewhat similar to the South side in some characteristics, including low median income, massive industrial lots, higher poverty rates, and limited economic opportunities. The West side has a Black and Hispanic population proportions intermediate between the two other neighborhoods. Figure 1 shows the locations of the three neighborhoods on the map of Chicago.

**TABLE 1 Data description**

| Category | Variable | Description | Code | | Percentage (%) | | |
|---|---|---|---|---|---|---|---|
| | | | | | North side | West side | South side |
| Socio-demographic variables | hhinc | Household income | 1 | Less than $15,000 | 2.50 | 5.39 | 11.03 |
| | | | 2 | $15,000 to $24,999 | 3.25 | 4.85 | 10.90 |
| | | | 3 | $25,000 to $29,999 | 2.53 | 3.36 | 7.58 |
| | | | 4 | $30,000 to $34,999 | 2.69 | 2.75 | 6.54 |
| | | | 5 | $35,000 to $49,999 | 7.37 | 8.50 | 12.40 |
| | | | 6 | $50,000 to $59,999 | 7.83 | 7.88 | 7.75 |
| | | | 7 | $60,000 to $74,999 | 11.81 | 9.01 | 11.17 |
| | | | 8 | $75,000 to $99,999 | 16.91 | 11.46 | 11.25 |
| | | | 9 | $100,000 to $149,999 | 21.83 | 22.71 | 14.07 |
| | | | 10 | $150,000 or more | 23.27 | 24.09 | 7.31 |
| | sex | Gender | 1 | Male | 46.25 | 41.92 | 37.28 |
| | | | 2 | Female | 53.75 | 58.08 | 62.72 |
| | race_x | Race | 0 | White | 85.86 | 72.12 | 36.16 |
| | | | 1 | Non-white | 14.14 | 27.88 | 63.84 |
| | hhveh_x | Household vehicle | 0 | No vehicle | 27.00 | 21.41 | 18.61 |
| | | | 1 | Have a vehicle | 73.00 | 78.59 | 81.39 |
| | hhsize_x | | 1 | 1 person | 28.30 | 22.10 | 23.90 |
| | | Household size | 2 | 2 persons | 43.70 | 49.40 | 28.96 |
| | | | 3 | 2+ persons | 28.00 | 28.50 | 47.14 |
| | age_x | Age | 1 | 16-29 years | 31.40 | 32.88 | 26.88 |
| | | | 2 | 30-44 years | 45.10 | 52.01 | 37.12 |
| | | | 3 | 45-64 years | 20.59 | 14.47 | 31.37 |
| | | | 4 | 64+ years | 2.91 | 0.65 | 4.63 |
| Trip characteristic | distance_x | Trip distance | 1 | Less than or equal to 0.25 mile | 12.14 | 11.65 | 8.57 |
| | | | 2 | Greater than 0.25 mile and less than equal to 0.5 mile | 12.95 | 12.51 | 8.46 |
| | | | 3 | Greater than 0.5 mile and less than equal to 1 mile | 12.71 | 12.95 | 11.74 |
| | | | 4 | Greater than 1 mile and less than equal to 2.5 miles | 16.19 | 20.51 | 17.57 |
| | | | 5 | Greater than 2.5 miles and less than equal to 5 miles | 15.94 | 22.06 | 15.74 |
| | | | 6 | Greater than 5 miles and less than equal to 10 miles | 18.89 | 9.95 | 18.29 |
| | | | 7 | Greater than 10 miles and less than equal to 25 miles | 8.77 | 7.67 | 16.78 |
| | | | 8 | Greater than 25 miles | 2.41 | 2.71 | 2.85 |
| | work_purp | Trip purpose | 0 | Non work related | 71.25 | 71.32 | 75.80 |
| | | | 1 | Work related | 28.75 | 28.68 | 24.20 |
| Mode choice | Car | Car mode | 0 | No | 60.62 | 55.70 | 40.24 |
| | | | 1 | Yes | 39.38 | 44.30 | 59.76 |
| | Public | Public transit mode (bus/train) | 0 | No | 72.18 | 74.94 | 77.03 |
| | | | 1 | Yes | 27.82 | 25.06 | 22.97 |
| | Walk | Walking mode | 0 | No | 67.20 | 69.37 | 82.73 |
| | | | 1 | Yes | 32.80 | 30.63 | 17.27 |
| N | | | | | 12245 | 2765 | 3653 |

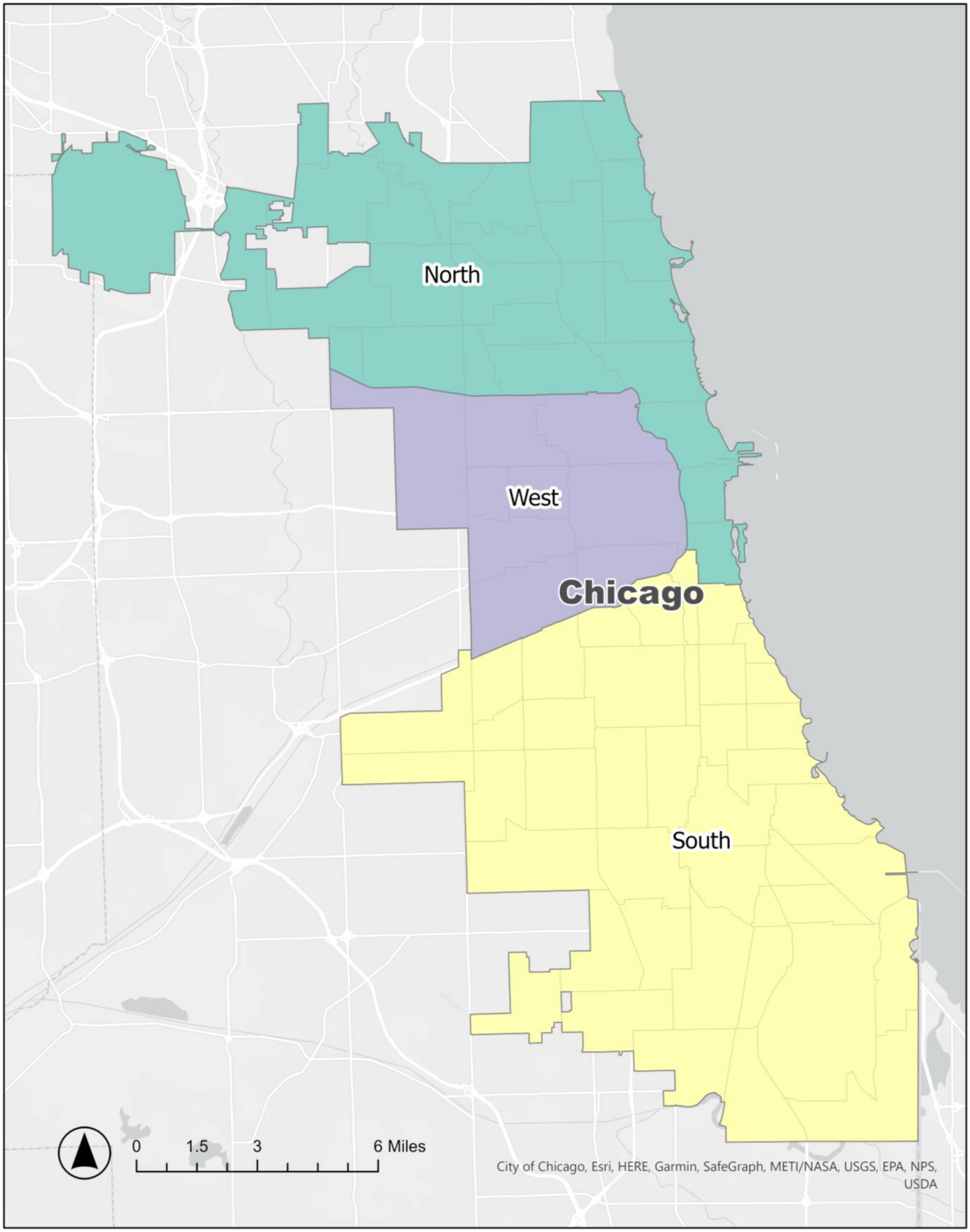

**Figure 1 Neighborhoods in Chicago**

**Basic concepts**

A directed acyclic graph (DAG) is denoted by $G = (V, E)$ where $V$ are vertices and $E$ are edges (48). DAGs are directed, meaning the edges have a direction, for example $V_1 \rightarrow V_2$. The node where an edge begins is called the parent node (like $V_1$ in $V_1 \rightarrow V_2$ which can be denoted as Pa($V_2$)), while the node where the edge points is called the child node (like $V_2$ in $V_1 \rightarrow V_2$). A path in a graph is defined as a sequence of adjacent edges. In a DAG, all the nodes preceding a node in a directed path are called its ancestors, while all the nodes succeeding it in a directed path are called its descendants. The Markov condition states that any node $X$ ($X$ Î $V$) is independent of its non-descendant nodes conditioned on its parent nodes (Pa($X$) $\subset V$) (48). Bayesian networks are used as a means to perform causal discovery and inference (48). For a given set of variables $V$, a Bayesian network could represent the joint probability distribution P($V$) through a DAG, assuming the Markov condition applies (48). A Bayesian network is considered a causal Bayesian network when its structure is considered causal, such that $V_1$ is considered a direct cause of $V_2$ in $V_1 \rightarrow V_2$ (48).

A path $V_1, \ldots, V_n$ in a DAG is said to be *blocked* by a set of nodes $Z$ (not consisting of $V_1$ or $V_n$) if: (a) there is a node $V_k$ in the path that is a non-collider (i.e., it is either $V_{k-1} \rightarrow V_k \rightarrow V_{k+1}$ or $V_{k-1} \leftarrow V_k \leftarrow V_{k+1}$ or $V_{k-1} \leftarrow V_k \rightarrow V_{k+1}$) and $V_k \in Z$ or (b) there is a node $V_k$ in the path that is a collider (i.e., $V_{k-1} \rightarrow V_k \leftarrow V_{k+1}$) and neither $V_k$ nor its descendants belong to $Z$ (10). If in a DAG ($G$), a set of nodes $Z$ blocks all the paths between two sets of nodes $A$ and $B$, given that $A$, $B$, and $Z$ are pairwise disjoint, then $A$ and $B$ are said to be *d-separated* by $Z$. This can be mathematically denoted as $A \perp_G B | Z$ (10). The global Markovian condition is satisfied in a DAG if for every pairwise disjoint $A, B, Z \subseteq V$, $A \perp_G B | Z$ implies $A \perp\!\!\!\perp B | Z$ (10).

There are two more relevant concepts that need to be discussed before moving on to the causal discover algorithms. These are the assumptions of faithfulness and causal sufficiency. The faithfulness assumption is the same as the global Markovian condition but reversed, such that $A \perp\!\!\!\perp B | Z$ implies $A \perp_G B | Z$ (10). As per the causal sufficiency assumption, all the common causes of any pair of variables in the observed data are also observed in the dataset (10).

**PC Algorithm**

There are several causal discovery algorithms suggested in the literature (10,49). In this article, we have chosen to use one of the most popular causal discovery algorithm, the PC algorithm (50). Note that finding the most suitable algorithm for mode choice modeling is beyond the scope of this article. Please refer to Chauhan et al. (36) for more on the topic. The PC algorithm uses conditional independence to extract causal relationships from observed data. With the assumptions of causal Markov, faithfulness, and no latent confounders, the PC algorithm suggests that two variables are directly causally related if and only if there exists no subsets of remaining variables conditioning on which the two variables are independent (49,50). As explained by Glymour et al. (49), the PC algorithm involves the following steps in sequence:

- The algorithm starts by assuming a complete undirect graph where every variable has an edge connecting to every other variable.
- The edge between a pair of variables (say $A$ and $B$) is eliminated if they are found to be conditionally independent i.e., $A \perp\!\!\!\perp B$.
- For each pair of variables (say $A$ and $B$) that have an edge between them, the edge is eliminated if $A \perp\!\!\!\perp B | C$ for any variable $C$ that has an edge connected to either $A$ or $B$.
- For each pair of variables (say $A$ and $B$) that have an edge between them, the edge is eliminated if $A \perp\!\!\!\perp B | \{C, D\}$ for any pair of variables $\{C, D\}$ that have edges both connected to either $A$ or $B$. Continue testing the independencies conditioned on subsets of variables of increasing size until there are no more left.
- For each triple (say $A$, $B$, and $C$) with an undirected edge between $A$ and $B$ and between $B$ and $C$ but not between $A$ and $C$ (i.e., $A – B – C$), be oriented to $A \rightarrow B \leftarrow C$ if $B$ was not in the set conditioning on which $A$ and $C$ became independent.

- For each triple (say *A*, *B*, and *C*) with a directed edge between *A* and *B* and undirected edge *B* and *C* but none between *A* and *C* (i.e., $A \rightarrow B - C$), $B - C$ should be oriented to $B \rightarrow C$ as per orientation propagation.

There may be other orientation propagation rules as well (49). An edge may remain undirected if none of the orientation rules of the PC algorithm apply (49). The Pycausal Python library was used to run the PC algorithm in this study (51).

**Double Machine**

There are several methods to estimate causal effects. In this study, we use the causal inference called DML (52). Assume a treatment variable *T* that has a causal effect on an outcome variable *O*, and a covariate variable *Z* that causes both *T* and *O*, DML performs three main steps to estimate the causal effect of *T* on *O*: (a) two ML models are estimated – first to predict *O* by *Z* and second to predict *T* by *Z*; (b) the residuals from both the models are computed; (c) the residual from the model predicting *O* by *Z* is regressed on the residual from the model predicting *T* by *Z* (52). Please refer to Chernozhukov (52) for details.

In this study, DML involves two Gradient Boosting Regression models to model the outcome and the treatment, and a Lasso Linear Regression model to estimate the residuals. The DoWhy Python library (53) was used to estimate the causal effects based on the causal graphs obtained from the PC algorithm.

**Domain Knowledge**

Inputting some domain knowledge to the causal discovery algorithms improves the model results (38,54). Therefore, we added some basic domain knowledge to the algorithms. However, we kept it to the very minimum not to bias the model results. The following rules were added to the algorithm:

- Since the goal of causal modeling is mode choice. The mode choice variables – car, public, and walk – were considered as the target variables, and therefore they cannot cause any other variables.
- Gender, race, and age were considered to be exogenous variables; i.e., these cannot be caused by any of the other variables.
- It is assumed that gender, age, and household income cannot cause household size.
- It is also assumed that vehicle ownership cannot cause household income. This may be considered arguable; however, it was hypothesized that household income causing vehicle ownership is more likely than the opposite.

**Artificial Neural Networks**

To develop the ANN models for the three neighborhoods, we had to adopt a pragmatic hyperparameter tuning approach. Since the dataset is relatively small, hyperparameter tuning is performed by combining the data for all three neighborhoods. The data are split into two sets, 80% training and 20% testing with stratified sampling. The training data are further split into three sets, 72% training, 18% validation, and 10% testing to tune the hyperparameters. Keras_tuner is employed to find the best hyperparameters. After trying a different number of layers and neurons with different activation functions, the highest accuracy with the simplest model structure is adopted. In our case, it is as follows: one hidden layer with 28 neurons, the Scaled Exponential Linear Unit (SELU) activation function, and a learning rate of 0.0005. Subsequently, 5-fold cross validation is performed, and we find that all five models have a consistent accuracy of ~75%. We then train the model one more time on the combined train and validation data, and we test it on the test data to validate the choice of hyperparameters. This step gave us an acceptable accuracy of 74%.

Finally, the model is trained separately on the original 80% training data for each neighborhood and tested one final time on the original testing data for each neighborhood. We note that the testing data was kept separate while tuning the model hyperparameters to avoid data leakage. The accuracies were calculated for each neighborhood using their respective final test sets.

The accuracy of the models was calculated as:

$$Accuracy = \frac{Total\ number\ of\ correct\ preditions}{Total\ of\ predictions} \times 100 \quad (1)$$

No scaling was applied since all the features are either nominal binary or ordinal categorical. Data were oversampled using Synthetic Minority Over-sampling Technique (SMOTE) (4,55) to balance the fewer samples for the public transit and walk modes compared to the car mode. The Keras and SHAP Python libraries were used for the analysis.

SHAP (SHapely Additive exPlanations) values improve the interpretability of ML models (25). These quantifies the contribution of each feature to the prediction of the model by evaluating the marginal contribution of that feature (56,57). The formulation of SHAP value can be expressed as follows (58):

$$\emptyset_i = \sum_{S \subseteq F\{i\}} \frac{|S|!(|F|-|S|-1)!}{|F|!} [f_{S\cup\{i\}}(x_{S\cup\{i\}}) - f_S(x_S)] \quad (2)$$

where, $|F|$ is the set of all combinations of features, $S$ is a subset of $|F|$ excluding feature $i$, $|S|$ is the size of subset $S$, and $f_{S\cup\{i\}}(x_{S\cup\{i\}}) - f_S(x_S)]$ is the added value after including feature $i$ with $S$.

## RESULTS

### Causal Models

Figure 2 shows the causal graphs obtained from the PC algorithm. There was one edge in each of the three graphs that the algorithm was not able to direct. These were work_prup – distance_x in the North and South sides, and hhsize – hhinc in the West side. These were oriented as work_prup → distance_x and hhsize → hhinc based on domain knowledge. Table 2 shows the total causal effects obtained from DML. These values are the average treatment effects (ATE), meaning the average difference in the pair of potential outcomes averaged over the entire dataset. These causal effects can be interpreted as the average change in outcome caused by the treatment. A positive value suggests an increase in outcome, while the negative value suggests a decrease.

The causal graphs present a visual representation of the flow of causality over the decision variables and shed light on mode choice decision making processes. Figure 2 shows that vehicle ownership and trip distance are the two direct two causes of choosing any mode in any neighborhood. It is only in the South side that race is also a direct cause of walking (discussed below). The direct causes of vehicle ownership are household size (in all three neighborhoods), household income (in the West and South sides), and race (in the South side). The direct cause of trip distance, in the North and South sides, is work related purpose of trip.

The total causal effects on mode choice also provide meaningful Insights. In the North side, the causal impact on choosing car comes from vehicle ownership (0.41), household size (0.09), trip distance (0.08), and work-related purpose of trip (-0.04). Further, the causal impact on choosing public transit comes from household vehicle (-0.31), work-related purpose of trip (0.13), trip distance (0.10), and household size (-0.06). Finally, the causal impact on choosing to walk comes from trip distance (-0.17), vehicle ownership (-0.10), work-related purpose of trip (-0.09), and household size (-0.06).

In the West side, the causal impact on choosing car comes from vehicle ownership (0.49), trip distance (0.08), white race (-0.06), and household income (0.02). Further, the causal impact on choosing public transit comes from vehicle ownership (-0.38), trip distance (0.07), white race (0.06), household income (-0.02). Finally, the causal impact on choosing to walk comes from trip distance (-0.16), vehicle ownership (-0.11), white race (-0.04), household size (-0.03), and household income (0.01).

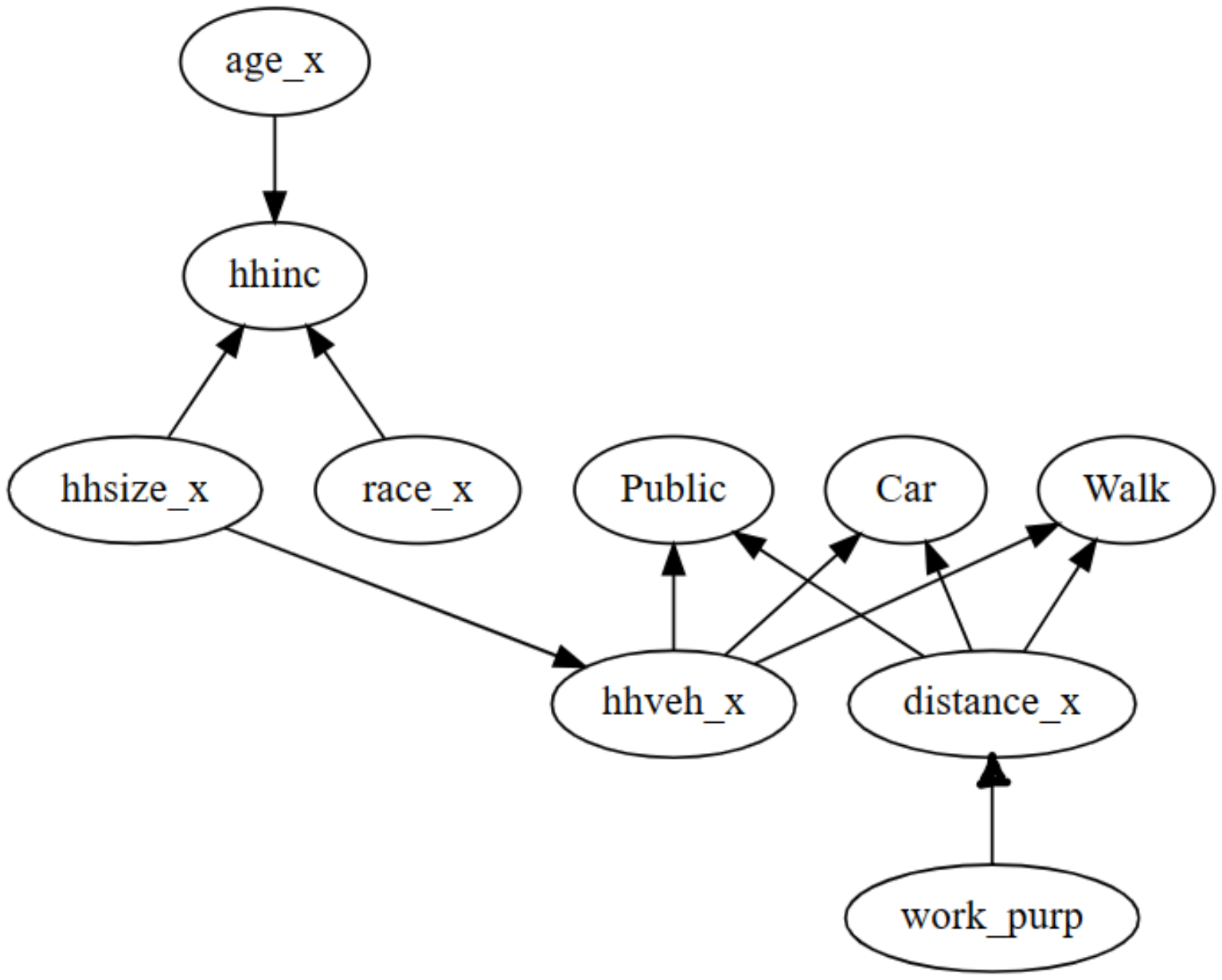

**(A)**

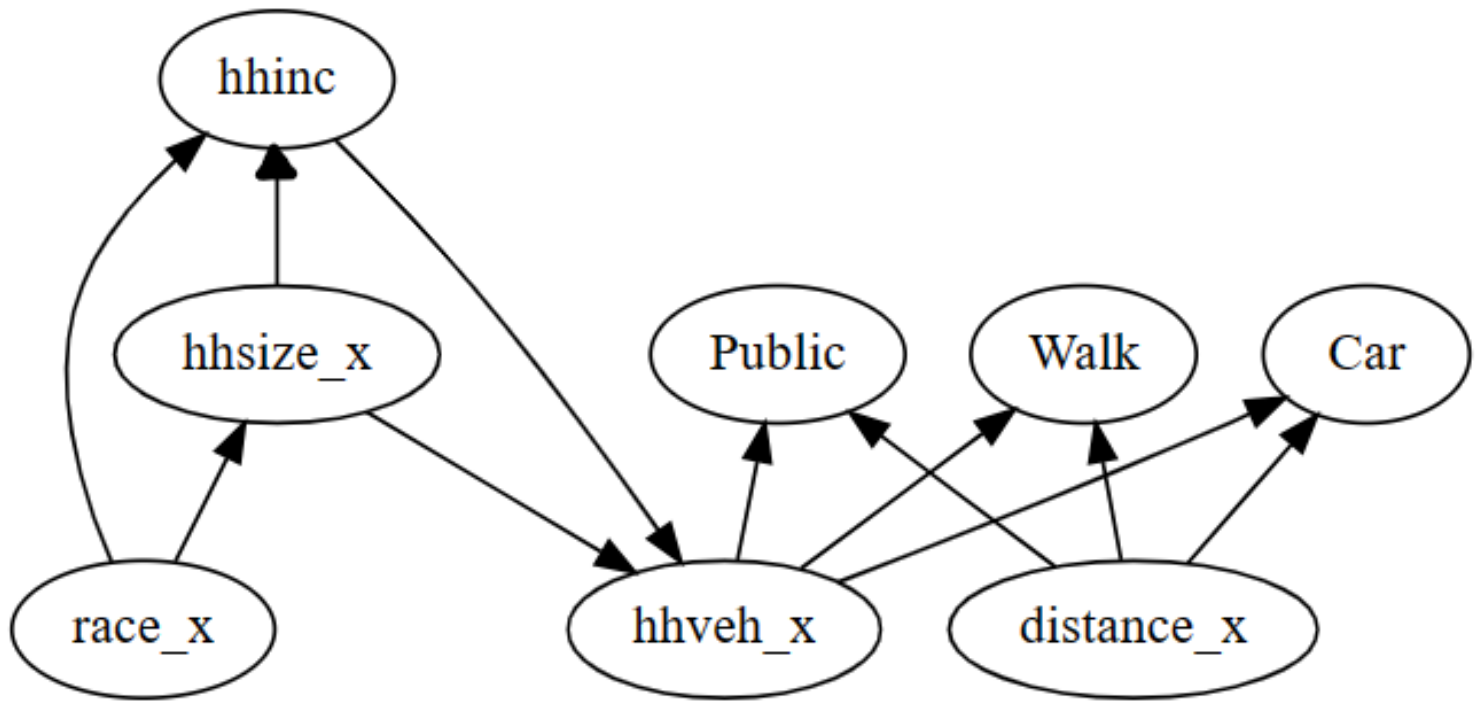

**(B)**

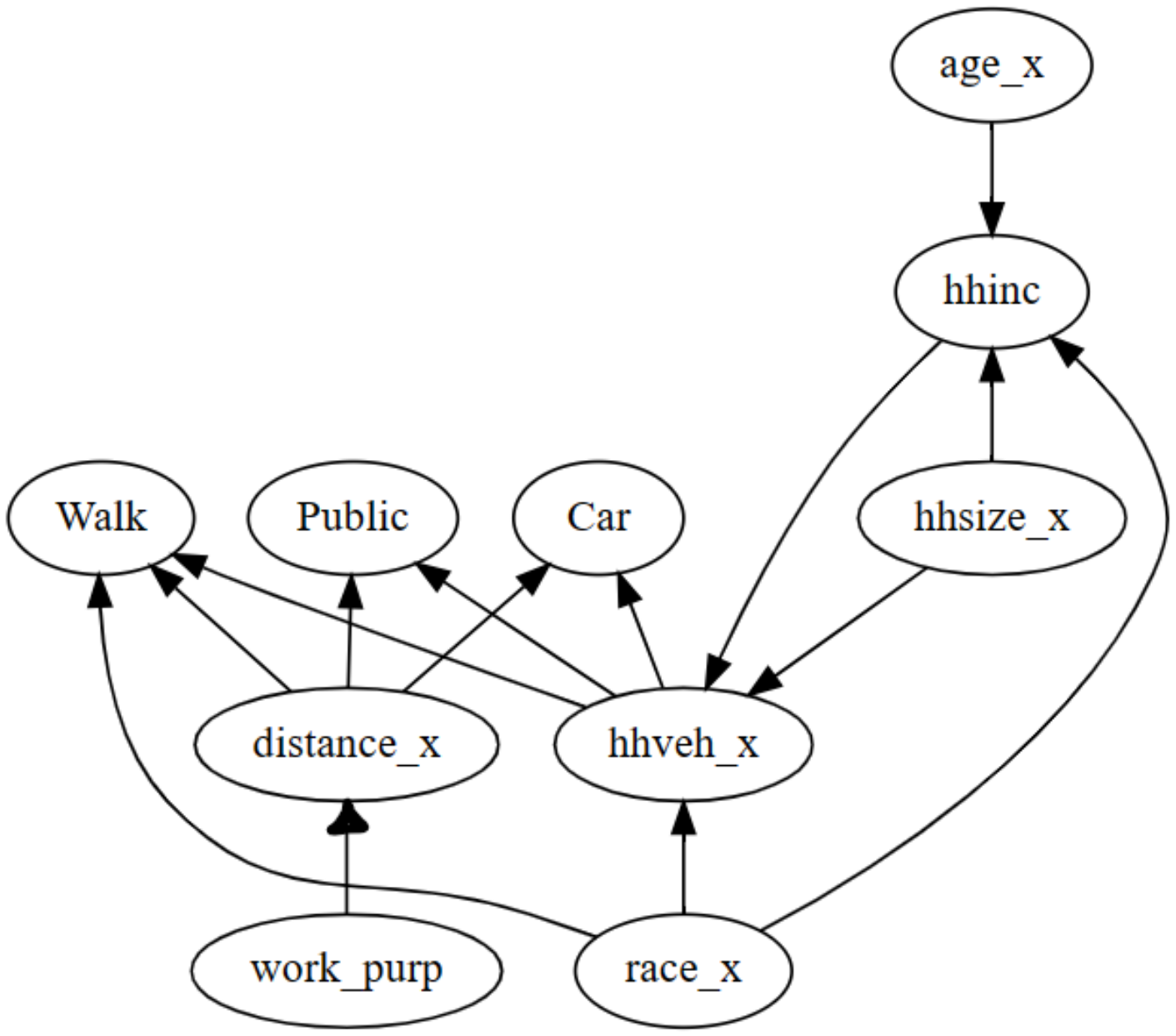

**(C)**

**Figure 2 Causal graphs for the (A) North side, (B) West side, and (C) South side of Chicago**

**TABLE 2 Total causal effects for the different neighborhoods**

| | Cause | Effect | | | | | | | | | | |
|---|---|---|---|---|---|---|---|---|---|---|---|---|
| Neighborhood | | hhinc | race_x | hhveh_x | hhsize_x | age_x | distance_x | work_purp | sex | Car | Public | Walk |
| North side | hhinc | - | 0.00 | 0.00 | 0.00 | 0.00 | 0.00 | 0.00 | 0.00 | 0.00 | 0.00 | 0.00 |
| | race_x | -1.09 | - | 0.00 | 0.00 | 0.00 | 0.00 | 0.00 | 0.00 | 0.00 | 0.00 | 0.00 |
| | hhveh_x | 0.00 | 0.00 | - | 0.00 | 0.00 | 0.00 | 0.00 | 0.00 | 0.41 | -0.31 | -0.10 |
| | hhsize_x | 0.97 | 0.00 | 0.20 | - | 0.00 | 0.00 | 0.00 | 0.00 | 0.09 | -0.06 | -0.06 |
| | age_x | 0.53 | 0.00 | 0.00 | 0.00 | - | 0.00 | 0.00 | 0.00 | 0.00 | 0.00 | 0.00 |
| | distance_x | 0.00 | 0.00 | 0.00 | 0.00 | 0.00 | - | 0.00 | 0.00 | 0.08 | 0.10 | -0.17 |
| | work_purp | 0.00 | 0.00 | 0.00 | 0.00 | 0.00 | 0.58 | - | 0.00 | -0.04 | 0.13 | -0.09 |
| | Sex | 0.00 | 0.00 | 0.00 | 0.00 | 0.00 | 0.00 | 0.00 | - | 0.00 | 0.00 | 0.00 |
| West side | hhinc | - | 0.00 | 0.04 | 0.00 | 0.00 | 0.00 | 0.00 | 0.00 | 0.02 | -0.02 | 0.01 |
| | race_x | -2.48 | - | -0.07 | 0.10 | 0.00 | 0.00 | 0.00 | 0.00 | -0.06 | 0.06 | -0.04 |
| | hhveh_x | 0.00 | 0.00 | - | 0.00 | 0.00 | 0.00 | 0.00 | 0.00 | 0.49 | -0.38 | -0.11 |
| | hhsize_x | 0.70 | 0.00 | 0.14 | - | 0.00 | 0.00 | 0.00 | 0.00 | 0.00 | 0.00 | -0.03 |
| | age_x | 0.00 | 0.00 | 0.00 | 0.00 | - | 0.00 | 0.00 | 0.00 | 0.00 | 0.00 | 0.00 |
| | distance_x | 0.00 | 0.00 | 0.00 | 0.00 | 0.00 | - | 0.00 | 0.00 | 0.08 | 0.07 | -0.16 |
| | work_purp | 0.00 | 0.00 | 0.00 | 0.00 | 0.00 | 0.00 | - | 0.00 | 0.00 | 0.00 | 0.00 |
| | sex | 0.00 | 0.00 | 0.00 | 0.00 | 0.00 | 0.00 | 0.00 | - | 0.00 | 0.00 | 0.00 |
| South side | hhinc | - | 0.00 | 0.00 | 0.00 | 0.00 | 0.00 | 0.00 | 0.00 | 0.02 | -0.02 | 0.00 |
| | race_x | -1.82 | - | -0.08 | 0.00 | 0.00 | 0.00 | 0.00 | 0.00 | -0.05 | 0.08 | -0.05 |
| | hhveh_x | 0.00 | 0.00 | - | 0.00 | 0.00 | 0.00 | 0.00 | 0.00 | 0.56 | -0.47 | -0.11 |
| | hhsize_x | 0.60 | 0.00 | 0.11 | - | 0.00 | 0.00 | 0.00 | 0.00 | 0.05 | -0.03 | -0.01 |
| | age_x | 0.57 | 0.00 | 0.06 | 0.00 | - | 0.00 | 0.00 | 0.00 | 0.07 | -0.05 | -0.02 |
| | distance_x | 0.00 | 0.00 | 0.00 | 0.00 | 0.00 | - | 0.00 | 0.00 | 0.06 | 0.05 | -0.10 |
| | work_purp | 0.00 | 0.00 | 0.00 | 0.00 | 0.00 | 0.57 | - | 0.00 | -0.04 | 0.09 | -0.02 |
| | sex | 0.00 | 0.00 | 0.00 | 0.00 | 0.00 | 0.00 | 0.00 | - | 0.00 | 0.00 | 0.00 |

In the South side, the causal effects on mode choice are spread out over several causes. All the variables which were suspected to have causal effect on mode choice (household income, white race, vehicle ownership, household size, age, trip distance, and work-related trip purpose) had a causal effect on all three modes. Although vehicle ownership stands out among all other variables due to its high causal impact on choosing a car and public transit (0.56 and -0.47 respectively).

It is interesting to note that race has no causal effects on choosing any modes in the North side, while it does on every mode in the West and the South side. Race is also a direct cause of household income, and it has a very high causal effect on household income which is even more pronounced in the West and the South side. More generally, the inclusion of race as an explanatory variable can be controversial and is discussed below. Another key finding is that household size was found to be a direct cause of vehicle in all three neighborhoods.

## Artificial neural network models

The ANN models when applied to each neighborhood achieved an accuracy of 75.46%, 74.68%, and 72.50% for the North side, the West side, and the South side, respectively.

Figure 3 shows the SHAP values of the different variables on the model output. These values are based on prediction and are average impact of the variables on model prediction. For each neighborhood, the variables are ordered based on their magnitude of influence (with the largest at the top). In all three neighborhoods, the most important variables are trip distance and vehicle ownership. Notably, race has a more significant role in the South sides than in the North and West side. The variables importance also varies with the type of mode. In all neighborhoods, trip distance has the biggest influence on walking. Vehicle ownership has the highest impact on choosing the car mode in North and South while, for West, it is distance. Household size plays an important role in car choice in the North side while, in the West and South, it is household income. Further, the biggest impact on choosing public transit is from trip distance in the North and West sides and vehicle ownership in the South side. However, we note that for all three neighborhoods, the models performed worse for public transport than car and walk mode. This can be indicative of the fact that demographic attributes/ user specific variables do not always behave consistently in predicting mode choice for public transport (59).

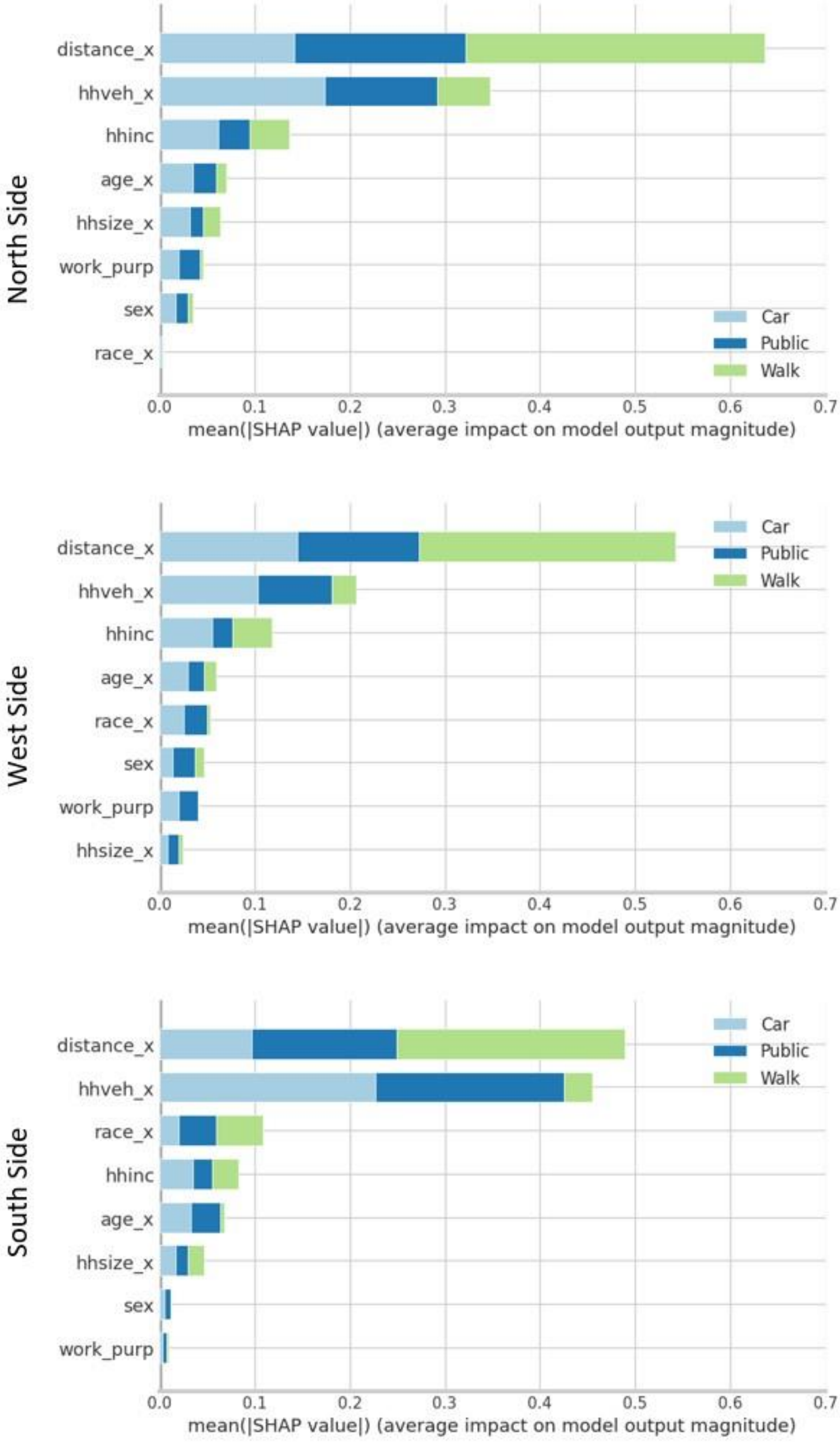

**Figure 3 Average SHAP values for the three ANN models**

## DISCUSSION

In this section, we comment on the two modeling approaches for mode choice modeling:

i. *Causal modeling and predictive modeling are two distinct modeling approaches*. As mentioned earlier, the goal of causal modeling is to understand the causal relations while that of predictive modeling is to predict the future (1). Both are suitable for their own specific applications. When selecting between the two, the modeler needs to contemplate the purpose of modeling. Travel mode choice modeling may be one of the unique avenues where both approaches can find their applications. Causal modeling could be helpful for knowing where to intervene to bring a mode shift, while predictive modeling could aid in estimating the need for infrastructure in the future.

ii. *Results from the causal model and the predictive model show some similarities*. Despite their fundamental differences, the results from both approaches exhibit some notable similarities. For example, the causal models suggest that trip distance and vehicle ownership are the direct causes of choosing any mode in all neighborhoods (**Figure 2**). These two variables also have the highest SHAP values in all neighborhoods (**Figure 3**). Similarly, causal models found race to have no causal effect on choosing any mode in the North side, unlike the other two neighborhoods. This is reflected in the SHAP values where race is at the bottom of the list in the North side but is important in the West and South sides. It is unclear whether these similarities are only a coincidence or if they point to something meaningful about the two approaches and/or data. Further analysis is needed on this point.

iii. *Causal models are built on some noteworthy assumptions*. The causal discovery and inference techniques are based on certain assumptions. For example, the methods used in this study (the PC algorithm and DML (with a Lasso Linear Regression model to estimate residuals) assume the absence of any unobserved confounders and linear variation in causal effects, respectively. Therefore, the true causal graph for mode choice may have more variables than those shown in Figure 2 and/or may have non-linear causal effects. Additionally, the variables in a causal model may be a proxy. In particular, the variable race is hypothesized to be a proxy for complex social, political, economic, and cultural factors as explained in Chauhan et al. (36). Researchers must be careful in picking the causal techniques and be mindful of their assumptions and the data used in the study.

iv. *Causal models can guide interventions, but predictive models may not*. Despite the few assumptions and limitations associated with the causal discovery and inference techniques, they are truly causal models. As such, they can extract the causal structure and estimate the causal effect from observed data. Being based on causation, they could suggest where to (and where not to) intervene to bring about a desired change. For example, intervening on any variable that has zero causal effect on choosing a mode will not lead to a change in the selection of that mode. The SHAP values from the predictive models cannot guide us about interventions since these values are based on the impact of a variable on prediction, which does not necessarily indicate a causal relation between the variable and the predicted variable. For instance, the variable sex does not even appear in the causal graph; however, it still has a non-negligible SHAP value.

v. *Both causal and predictive models may be capturing regional variations in travel behavior*. Firstly, the causal graphs and SHAP values have remained moderately similar across the neighborhoods. This makes sense since all three datasets are from the same city with some geographic variation in travel behavior. This may also be indicative of the robustness of both models. Secondly, models show that walking in the South side is different from the North side. The causal graph suggests that race is a direct cause of walking in the South side (unlike anywhere else). The SHAP values of race for predicting walking is much bigger in the South side as well. Here again, race may be a proxy for complex factors, possibly around safety.

vi. *Causal models have much unexplored potential in mode choice modeling*. Mode choice modeling has been mostly conducted through predictive modeling. Causal discovery and inference techniques have rarely been applied and show much promise. Causal modeling could create new avenues to study mode choice and reshape our understanding of the mode choice behavior.

**CONCLUSION AND FUTURE WORK**

As of this writing, travel mode choice studies have been mostly dominated by correlation-based models, including both statistical and ML models. Studies on causal modeling of mode choice remain limited.

New causal discovery and inference techniques allow for the estimation of data-driven causal models. A combination of the PC algorithm and DML is used in this article to estimate the causal relationships to model mode choice in three Chicago neighborhoods. The result is a graphical representation of the flow of causality in the decision-making process as well as the quantitative estimation of the causal effects among the variables. We find that the two direct causes of choosing any mode (car, public transit, or walk) in any of the three neighborhoods are trip distance and vehicle ownership. Race was also found to be a direct cause of walking in the South side. The estimated causal effects present the strength of any direct or indirect causal effects in the mode choice process.
For comparison, three ANN models were also estimated and performed relatively well (over 70% accuracy). The SHAP values from these models provided insights on the importance of each variable. Even though causal modeling and predictive modeling are fundamentally different, they found many similar results. Both models seem to capture some geographic variation in travel behavior in Chicago. Overall, both modeling approaches have their usefulness in mode choice modeling.

The study found an important potential for causal discovery and inference to model travel mode choice. The causal modeling of mode choice reveals the underlying causal relations from the data and could advance our understanding of the mode choice behavior. More research is needed to realize the full potential of these models, particularly about (a) the choice of causal discovery and inference techniques most suitable for mode choice studies, (b) investigation of the causal model results developed on different datasets with more variables, and (c) using causal models to make predictions and to compare the accuracy with the predictive models.

**AUTHOR CONTRIBUTIONS**

The authors confirm contribution to the paper as follows: study conception and design: RSC, SD; data collection: not applicable; analysis and interpretation of results: RSC, US, AR, SD; draft manuscript preparation: RSC, US, AR, SD. All authors reviewed the results and approved the final version of the manuscript.